\documentclass{elsart}
\usepackage{epsfig}
\usepackage{graphicx}
\begin{document}
\runauthor{Marina and JC}
\begin{frontmatter}
\title{Study of the spatial resolution achievable with the BTeV 
pixel sensors}
\author{Marina Artuso and Jianchun Wang}
\address{Department of Physics, Syracuse University, Syracuse, NY 13244}
\begin{abstract}
A Monte Carlo simulation has been developed to predict the spatial 
resolution of silicon pixel detectors. The results discussed in this paper 
focus on the
unit cell geometry of 50 $\mu$m $\times$ 400 $\mu$m, as chosen for BTeV. 
Effects 
taken into account include energy deposition fluctuations
along the charged particle path, diffusion, magnetic field,
and response of the front end electronics. 
We compare our predictions with
measurements from a recent test beam study performed at Fermilab.
\end{abstract} 
\end{frontmatter}

\section{Introduction}
BTeV is an experiment designed to explore heavy flavor phenomenology
thoroughly \cite{btevprop}, with particular
emphasis on mixing, $CP$ violation and rare and forbidden decays.  The
pixel
detector system is crucial to the experiment. We have
engaged in a thorough study of several different 
factors influencing the sensor performance, with particular emphasis on 
spatial resolution and detector occupancy. As the radiation 
dose is very intense at small distances from the beam axis, our
main emphasis is 
on $n^+np^+$ sensors \cite{atlas}, where the collected charge carriers 
are electrons.

We report the most important results on the expected performance of the
BTeV
baseline front end-sensor systems, as well as a comparison between our
predictions and results recently obtained in a test beam run that took place at
Fermilab, where we took extensive data using different prototype sensors and 
readout electronics \cite{gabriele}.

\section{Spatial resolution studies} 
We have
 investigated several factors affecting the spatial resolution, in
particular charge diffusion,
magnetic field, electronic noise, discriminator threshold and digitization 
resolution.

We have modeled the signal induced by minimum ionizing tracks traversing 
silicon 
using a charge straggling distribution function supported by experimental
data and a detailed theoretical model of the interactions responsible for
the energy loss in silicon \cite{bichsel}. The detector has been 
conceptually divided
into $\sim$ 30 $\mu$m thick slices to model fluctuations in the energy deposition 
along the charged particle track path. The most probable charge signal 
produced by a minimum ionizing particle 
in
the 280 $\mu$m p-stop sensors used in the test beam data is found to be  
24000 $e^-$ 
with a full width at half maximum of 10000 $e^-$, in good agreement with the
corresponding parameters of the measured distribution \cite{gabriele}.

Electrons and holes produced by the energy deposited by a traversing 
charged particle drift along the electric field 
lines ($\vec{E}$) in the detector.
The equations describing this drift motion are:
\begin{eqnarray}
\vec{J}_e = - q\rho _e\mu _e\vec{E}\\
\vec{J}_h = q\rho _h\mu _h\vec{E}
\end{eqnarray}
where $q$ is the magnitude of the electron charge, $\mu _e$ and $\mu _h$
are the mobility of electrons and
holes respectively, $\rho _e$ is the number of free electrons per unit volume,
 and $\rho _h$ is number of free holes per unit volume.

\begin{figure}[htbp]
\leavevmode
\begin{center}
\epsfysize=5 in
\epsffile{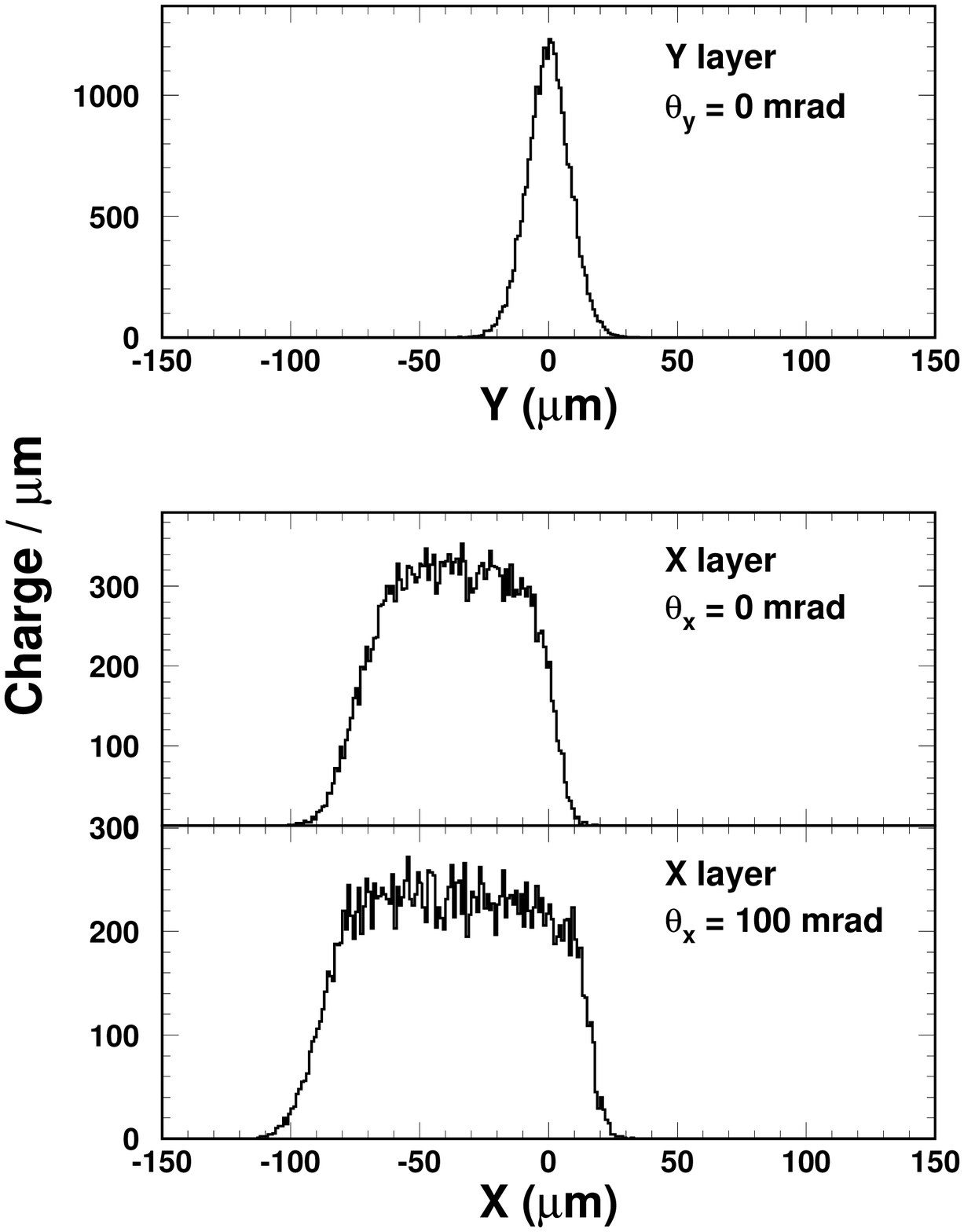}
\caption{\label{chdist} Collected charge spreading in a 280 $\mu$m silicon 
detector
a) produced by diffusion (the label Y layer identifies a pixel sensor 
oriented with the small pixel dimension parallel to the magnetic field);  
b) produced by the interplay of diffusion and the magnetic field for B=1.6T,
and c) produced by diffusion and magnetic field effects when the charged track 
is incident at an angle
of 100 mr in the bend plane (the label X layer identifies a pixel sensor 
oriented with the small pixel dimension perpendicular to the magnetic field).}
\end{center}
\end{figure}

The charge cloud spreads laterally due to diffusion. The 
parameters 
characterizing  the drift in the 
electric field ($\mu _h$, $\mu _e$) are related to the parameter describing 
the diffusion of the charge cloud 
($D_h$,$D_e$) by the Einstein equation:
\begin{equation}
D_{h(e)}=\frac{kT}{q} \mu_{h(e)}
\end{equation}
where $D_h$ and $D_e$ are the diffusion coefficients and $kT$ is 
the product of the
Boltzmann constant and the absolute temperature of the silicon.
The average square deviation with respect to the trajectory of the collected 
charge without diffusion is $<\Delta r^2>=2D\Delta t$. In our study we have 
used $\mu _h =400~{\rm cm^2/Vs},\ \mu_e =1450~{\rm cm^2/Vs}$ \cite{heine}. 
A magnetic
field perpendicular to $\vec{E}$
produces a distortion of the collected charge distribution
parameterized
in terms of the Hall mobility $\mu _H$, proportional to the drift
mobility.
Finally, the charge-cloud is collected on more pixels when the incident track 
crosses
the detector at an angle, as the generation points of the electron-hole pairs 
spread
out along the track path. These various effects are illustrated in 
Fig.~\ref{chdist} as a function of the incident angle $\theta$.



\begin{figure}[htbp]
\begin{center}
\epsfysize=6 in
\epsffile{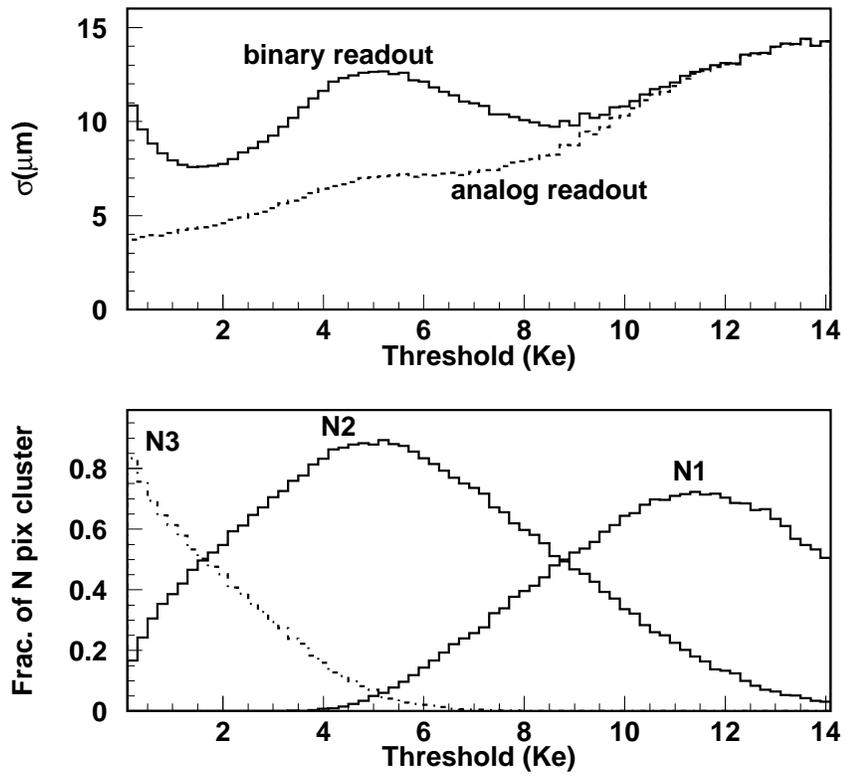}
\end{center}
\vspace{-1cm}
\caption{\label{resth} a) Spatial resolution in the reconstructed x 
coordinate as a function of
the threshold for a pixel size of
50 x 400 $\mu$m$^2$ and incidence angle of $\theta =300$ mr.
b) Percentage of events having N pixels hit as a function of the threshold
for the same configuration. No magnetic field is applied in this simulation.}
\end{figure}

 The discriminator threshold also influences the spatial
 resolution. In this study we have assumed that only 
the pixels having a signal above threshold are read out, and we have varied 
the threshold for analog and 
binary readout. Fig.~\ref{resth} (top) shows the effect of increasing the 
threshold for 
an incident angle $\Theta =300$ mr, for analog and digital readout 
respectively.  The bottom plot shows the fraction of events having $N$ 
pixels hit for a given 
threshold. 
For instance, for a threshold of 1000 electrons, about 60\% of the events 
have 3 pixels hit, and about 40\% have 2 pixels hit. The resolution 
achievable for binary readout shows a characteristic 
oscillatory
behavior as we change the threshold. As the digital clustering algorithm
exploits the information provided by the number of pixels in a cluster, its
 accuracy is best when there is an almost equal population in two different
 cluster sizes: the smaller cluster corresponding to a track incident in a
 pixel center and the bigger cluster corresponding to incidence close to the
boundary between two pixels.
In the analog readout case, the accuracy of any position reconstruction
algorithm is degraded as the threshold increases.

A pixel detector has the potential for being a very low 
noise system, since the capacitive load at the input of the charge sensitive 
preamplifier is negligible. We have achieved \cite{abder}
100 $e^-$ noise or below on the test bench.
In order to take full advantage of
the low noise, the minimum threshold spread among channels
needs to be small. The measured spread in the devices used in
the test beam run is about 380 $e^-$ \cite{gabriele}. Prototypes of the more
advanced version of this design show that a threshold
spread of 180 $e^-$ or better is achievable. Therefore noise and 
threshold spread figures
are not a limiting factor in the detector performance.

\begin{figure}[htbp]
\begin{center}
\epsfysize=5 in
\epsffile{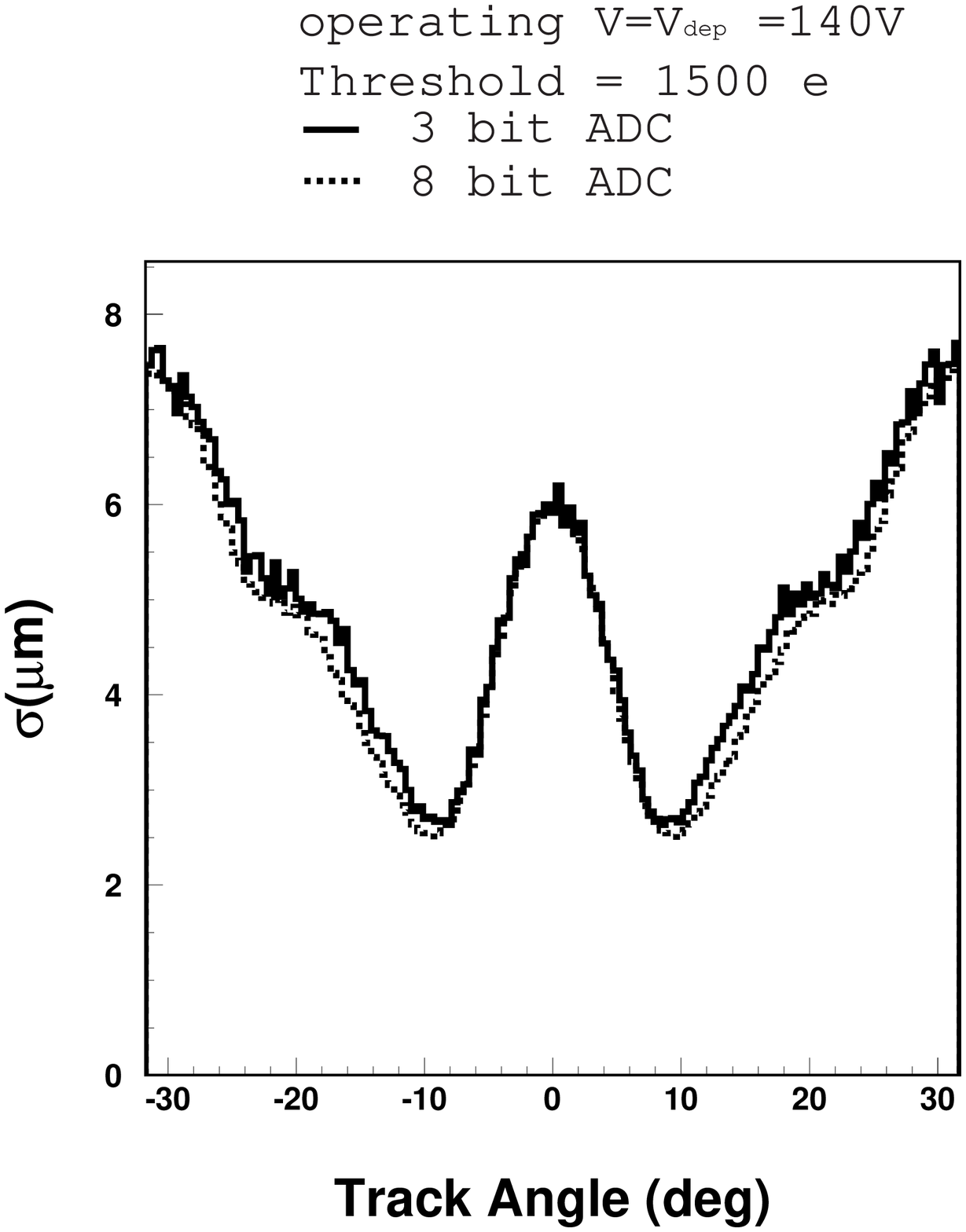}
\end{center}
\vspace{-1cm}
\caption{\label{fpix2} Sensitivity of the 
spatial resolution in the reconstructed x 
coordinate to the digitization resolution. A logarithmic 3 bit 
ADC (FPIX2) is compared to a linear 8 bit ADC. The threshold is
assumed to be 1500 $e^-$, with a dispersion of 200 $e^-$. The maximum
dynamic
range is assumed to be 20000 $e^-$. The sensor is biased at the depletion
voltage (assumed to be equal to 140 V) and has a thickness of 250 $\mu$m.}
\end{figure}

An analog readout is a preferred solution for several reasons, including 
more effective monitoring of the stability of the detector properties and 
improvement in the spatial resolution. We need to know how many
bits are required because the analogue circuit must fit within the small
pixel cell boundary. Furthermore, we need to extract the digitized
information very quickly.
 We have thus
investigated several different combinations of ADC transfer functions, varying
both the dynamic range and the digitization 
accuracy. Fig.~\ref{fpix2} shows an example where the expected performance of
our analog front end electronics combined with a logarithmic 3 bit flash ADC is
shown. As a comparison, the resolution  expected from a 8 bit linear ADC is 
included. Note that the difference is not too dramatic. This
supports our conclusion that the digitization resolution of 3 bits, chosen 
for our front end electronics, FPIX2, is optimal for our application.

\begin{figure}[htbp]
\begin{center}
\epsfysize=4 in
\epsffile{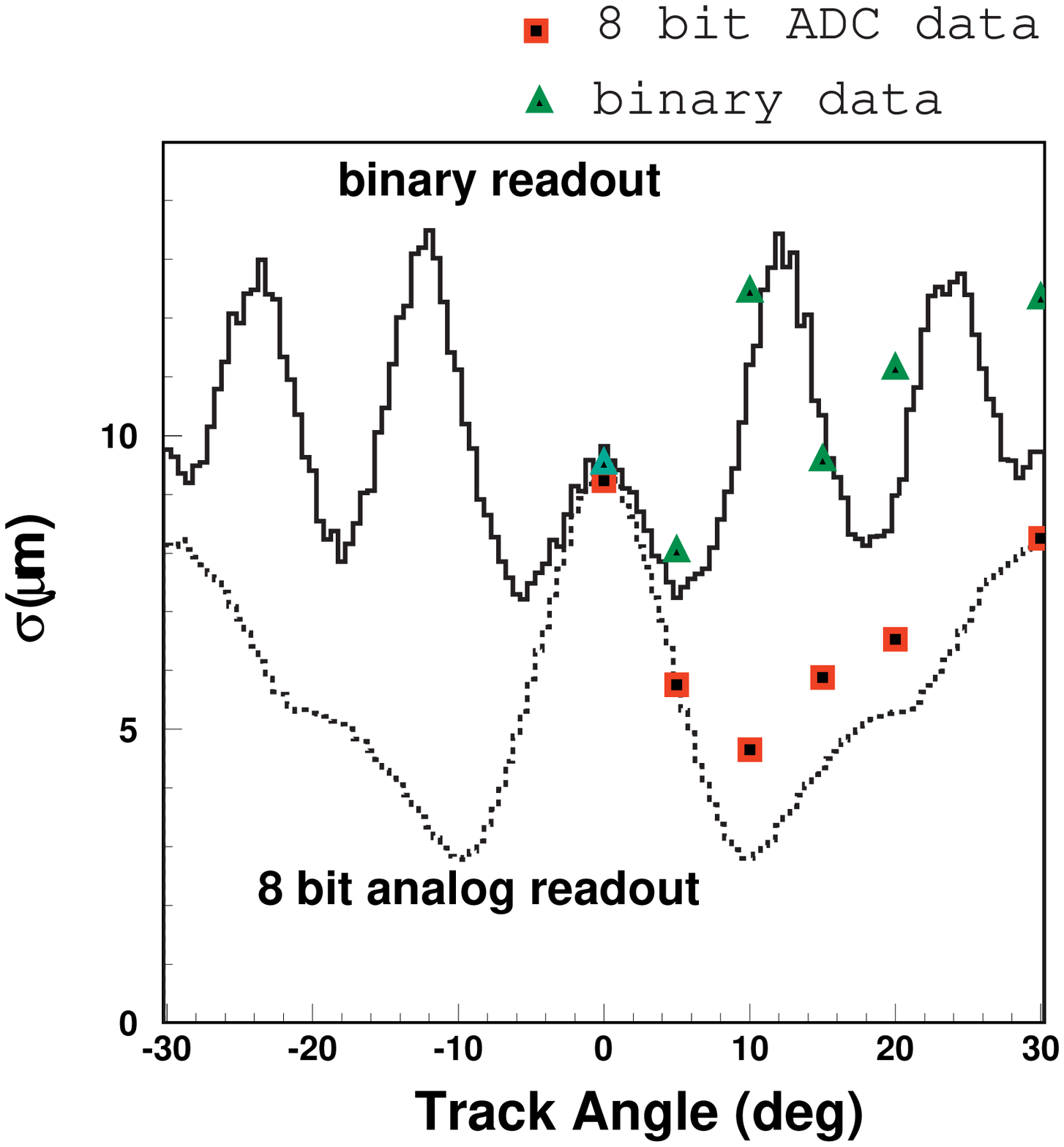}
\epsfysize=4 in
\vspace{1cm}
\epsffile{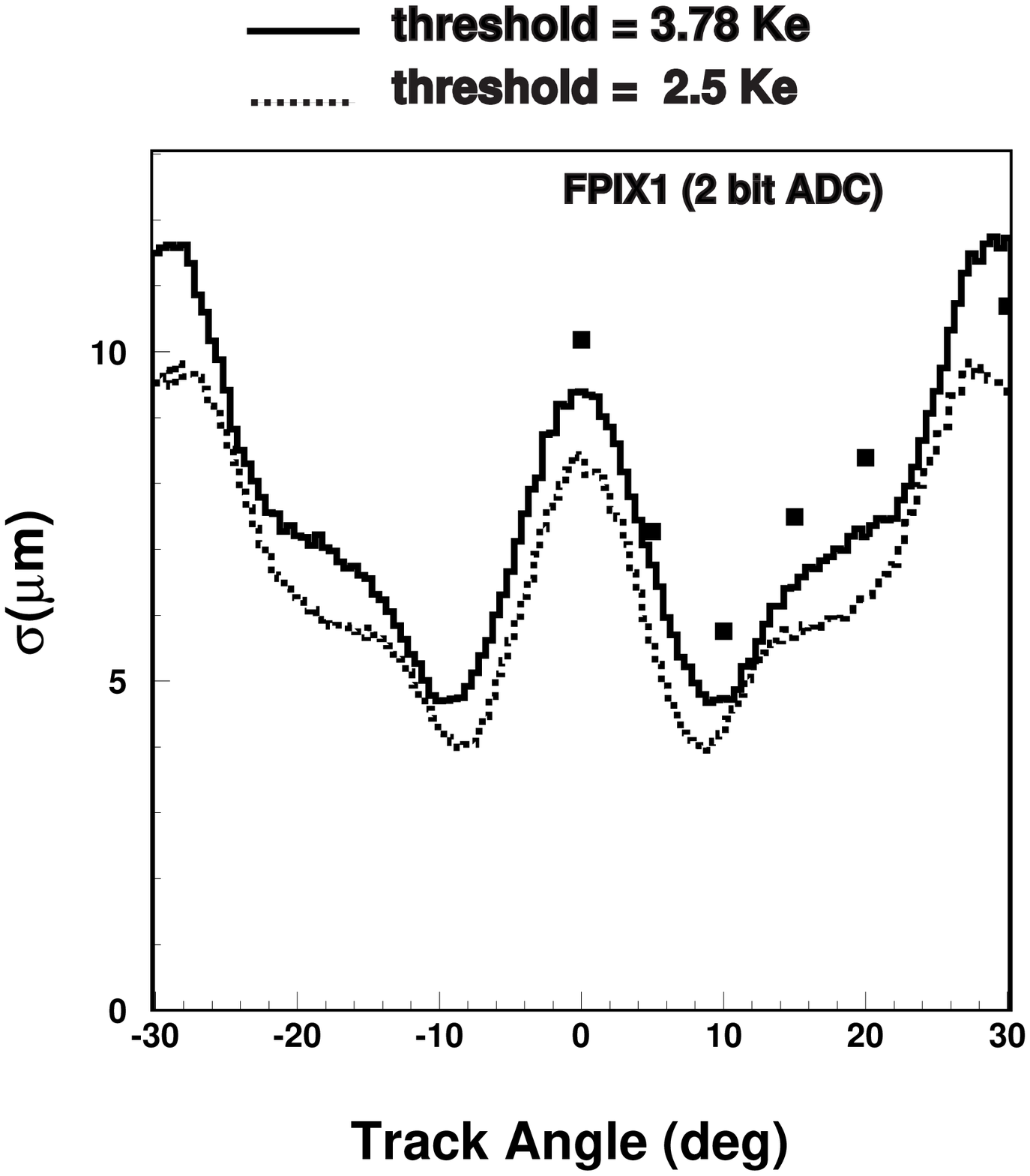}
\end{center}
\vspace{-1cm}
\caption{\label{resmix} Comparison between Monte Carlo predictions (curves)
and test beam data (points); top: FPIX0 data (8 bit ADC external to front end
chip, the curve labeled 'binary' corresponds to the same data analyzed using
only the pixel over threshold information; bottom: FPIX1, (2 bit ADC).}
\end{figure}

In order for this simulation tool to be effective, its accuracy 
must be checked with experimental data. We have done a systematic study of
the performance 
expected from various sensor and 
readout electronics combinations used in the recently completed test beam run.
Details on the data taken and the analysis procedure
are given elsewhere in these proceedings \cite{gabriele}. The comparison 
between
predicted and measured resolution is
shown in Fig.~\ref{resmix} for two different digitization accuracies (8 bit
and 2 bit ADCs) as well as binary readout. The data for the
two ADCs tested were  taken with different thresholds. Fig. \ref{resmix}
includes a simulation of the 2 bit ADC case with the lower threshold, showing
that the higher threshold used with the coarser ADC resolution is a dominant
effect, in particular at small incidence angles. 
The 
agreement with the data is very good, especially if we
take into account that factors like imperfect alignment, track projection
errors
and angular resolution have not yet been included. 

\section{Conclusions}
The Monte Carlo simulation algorithm discussed in this paper has been a key
element in our optimization of the baseline design of the BTeV pixel detector
system. The good agreement with recent test beam data gives us confidence that
the most important factors affecting the pixel performance are modeled
accurately and that this baseline design will be an excellent tool to achieve
our physics goals. 

While the results presented here are specific to the BTeV experiment, the
algorithm developed is quite general and can be applied to other detector
systems.

\section{Acknowledgements}
We thank D. Christian, S. Kwan, R. Mountain and S. Stone for 
interesting discussions. MA thanks the organizers for a very
informative workshop, in a beautiful setting. This work was supported by the
U.S. National Science Foundation.  


\vfill
\break
%
%

\begin{thebibliography}{999}
\bibitem{btevprop} The BTeV Collaboration, {\em The BTeV Proposal} (2000);
http://www-btev.fnal.gov/public\_documents/btev\_proposal/.
\bibitem{atlas} R. Horisberger, {\em Nucl. Instr. Meth.} {\bf A384} (1996) 185. \bibitem{gabriele}G. Chiodini, {\em These proceedings.}
\bibitem{bichsel}H. Bichsel {\em Rev. Mod. Phys.} {\bf 60} (1988) 663.
\bibitem{heine} L.H.H. Scharfetter (RD19), {\it Active Pixel Detectors for
Large Hadron Colliders}, CERN Thesis, (1997). 
\bibitem{abder}A. Mekkaoui, {\em These proceedings.}

\end{thebibliography}
\end{document}